\documentclass[english, russian, aps]{revtex4}
\usepackage{babel}
\usepackage{graphicx}
\begin{document}
\begin{large}

\title{Regularization of   the vacuum fluctuation  energy in quantum electrodynamics
 }

\author{I.D.Feranchuk, S.I.Feranchuk}

\affiliation{Belarusian State University, Nezavisimosti Ave., 4, 220030,
Minsk, Belarus\\
E-mail: fer@open.by}

\begin{abstract}

It is shown that the total   energy of the vacuum fluctuations of the electron-positron and electromagnetic fields in the quantum electrodynamics can be equal to zero if the "bare" electron charge  is chosen by adequate way. The  value of "bare" electron charge calculated from this condition proves to be in a qualitative agreement with that one  which was found independently  from the solution of the "physical electron" problem in \cite{Sigma2007}.

\end{abstract}
\maketitle

\section{Introduction}

It is no doubt at present that the quantum electrodynamics (QED) is actually the
part of the general gauge theory \cite{Weinberg}.   At the same time QED considered by itself as the isolated system
remains the most successful quantum field model that allows one to calculate the observed
characteristics of the electromagnetic processes with a unique accuracy using only two parameters - mass $m$ and charge $e$ of the "physical" electron (for example, \cite{Akhiezer}, \cite{LandauQED}, \cite{Bjorken}).
However, there are two "dark spots" in the theoretical construction of QED which do not permit to consider this model as the mathematically perfect theory \cite{Dirac} ( \S 81), \cite{Feynman1966}.

One of them is referred to the renormalization procedure where   the connection between characteristics of the "bare" electron (mass $m_0$ and charge $e_0$) and  parameters $m$ and   $e$ is defined by the divergent integrals. The second one is the infinite vacuum energy both of electron-positron and electromagnetic fields which is considered as the zero point energy. In spite of this infinite vacuum energy does not effect on the calculated amplitudes of the electrodynamical processes its reality appeared in the Casimir effect \cite{Casimir} (see, however, \cite{Jaffe}).

It was shown in series of our publications \cite{Sigma2007} , \cite{electron2003}, \cite{electron2006}  that it is possible to renormalize   mass and charge with finite values and connection between characteristics of the "bare" and "physical" electron is defined by the formulas \cite{Sigma2007}$(\hbar = c = 1)$ :

\begin{eqnarray}
\label{1}
\alpha_0 = \frac{a_0^2}{\alpha} \approx \frac{12.47}{\alpha} \approx 1708; \ \alpha_0 = \frac{e_0^2}{4 \pi}; \ \alpha  = \frac{e^2}{4 \pi}; \nonumber\\
m_0 = m \frac{2 |a_0|}{T \alpha} \approx \frac{9.43}{\alpha} m \approx 1292 m.
\end{eqnarray}

Here $\alpha$ is the observable fine structure constant, $a_0,T$ are the constants calculated numerically from the equations that define   one-particle excitation of the electron-positron field in QED. It was also shown that the perturbation theory on the physical value $\alpha$ corresponds to the strong coupling series on the parameter $\alpha_0^{-1}$.

In the  present paper we calculate the ground state energy (vacuum energy) $E_ 0 (\alpha_0) $ of the QED Hamiltonian   and show that it turns to be zero at the coupling constant $\tilde{\alpha}_0$( $E_ 0 (\tilde{\alpha}_0)   = 0)$. Calculated value  $\tilde{\alpha}_0 \gg 1$ is in a qualitative agreement with the coupling constant  (\ref{1}).

\section{QED vacuum energy}

Let us consider the QED Hamiltonian in the Coulomb gauge \cite{Bjorken}:

\begin{eqnarray}
\label{2}
\hat H  = \int d \vec r \hat H(\vec r) = \int d \vec r \{ \hat \psi^+(\vec r)[\vec \alpha (    \hat{\vec{p}}  - e_0 \hat{\vec A}(\vec r)) + \beta m_0]\hat \psi(\vec r) + \nonumber\\ + \sum_{\vec k \lambda} \omega (\vec k) (c^+_{\vec k \lambda} c_{\vec k \lambda} + \frac{1}{2})   + \frac{\alpha_0}{2  } \int d \vec r' \frac{\hat \rho (\vec r )\hat \rho (\vec r' )}{|\vec r - \vec r'|}\}; \nonumber\\
\hat \rho (\vec r ) =   \frac{1}{2}[\hat \psi^+(\vec r)\hat \psi(\vec r) - \hat \psi(\vec r)\hat \psi^+(\vec r)].
\end{eqnarray}

It is supposed here that the field operators are given in the Schr\"odinger  representation, the electromagnetic field operator  and the spinor components of
the electron-positron operators being defined in the standard way \cite{Akhiezer}:

\begin{eqnarray}
\label{3}
\hat{\vec A}(\vec r)) = \sum_{\vec k \lambda} \frac{\vec e^{(\lambda)}}{\sqrt{2 k V}} [c_{\vec k \lambda} e^{i \vec k \vec r} + c^+_{\vec k \lambda}e^{- \vec k \vec r}]; \nonumber\\
\hat \psi_{\nu} (\vec{r},t) = \sum_{s} \int \frac{d\vec{p}}{(2\pi)^{3/2}} \{a_{\vec{p}s}
u_{\vec{p}s\nu} e^{i\vec{p}\vec{r} } +
b^+_{\vec{p}s} v_{-\vec{p}-s\nu} e^{-i\vec{p}\vec{r}}\},
\end{eqnarray}
\noindent where $V$ is the normalization volume.

Let us remind briefly our  consideration of the one particle excitation \cite{Sigma2007}. In the zero order of the conventional  perturbation theory (PT) the one electron ($1e$) and one positron ($1p$) excitations are defined by the following state vectors

\begin{eqnarray}
\label{4}  |\Phi^{(PT)}_{1e}> = a^+_{\vec{p} s} | 0; 0; 0 >; \quad  |\Phi^{(PT)}_{1p}> =
b^+_{\vec{p} s} | 0; 0; 0 >,
\end{eqnarray}
\noindent with $|\Phi_0 > \equiv | 0; 0; 0 >$ as the vacuum state vector.

In our approach the more complicated trial state vectors was used for one particle excitation. It was represented by the  wave packet corresponding to "physical" electron or positron:

\begin{eqnarray}
\label{5} |\Phi^{(0)}_1 >  =   \sum_s\int d \vec{q} \{
U_{\vec{q}s} a^+_{\vec{q} s} + V_{\vec{q}s} b^+_{\vec{q} s} \} | 0;
0;0>.
\end{eqnarray}

Here $U_{\vec q s}; V_{\vec q s}$ are the variational parameters. For the "physical" electron with the zero total momentum they should satisfy the additional conditions:

\begin{eqnarray}
\label{6} <\Phi^{(0)}_1|\hat{ \vec {P}}|\Phi^{(0)}_1> = \sum_{s}\ d \vec{q} \vec{q} [|U_{\vec{q}s}|^2 +
 |V_{\vec{q}s}|^2] = \vec{P} = 0;\nonumber\\
<\Phi^{(0)}_1| \Phi^{(0)}_1> = \sum_{s}\ d \vec{q} [|U_{qs}|^2 + |V_{qs}|^2] = 1.
\end{eqnarray}

\begin{eqnarray}
\label{7} <\Phi^{(0)}_1|\hat Q|\Phi^{(0)}_1> = e_0 \sum_{s}\ d \vec{q} [|V_{qs}|^2 - |U_{qs}|^2] = e.
\end{eqnarray}
\noindent with $e$ as the "physical" electron charge.

Then we have calculated the one particle excitation energy defined as:

\begin{eqnarray}
\label{8} E_1 (0) \simeq  E^{(0)}_1 (U_{qs}; V_{qs}) - E_0 =  <\Phi^{(0)}_1|\hat H |\Phi^{(0)}_1> - <\Phi_0|\hat H |\Phi_0>,
\end{eqnarray}

\noindent where the averages are calculated with the total Hamiltonian (\ref{2}).

The functions $U_{qs};
V_{qs}$ were found as the solutions of variational equations:

\begin{eqnarray}
\label{9} \frac{\partial E^{(0)}_1 (U_{qs}; V_{qs})}{\partial U_{qs}} = \frac{\partial
E^{(0)}_1 (U_{qs}; V_{qs}) }{\partial V_{qs}} =    0
\end{eqnarray}

\noindent with the additional conditions   (\ref{6}) - (\ref{7}).

Using the coordinate representation

\begin{eqnarray}
\label{10} \Psi_{\nu} (\vec r) = \int \frac{d \vec q}{(2\pi)^{3/2}} \sum_{s} U_{q s} u_{\vec{q} s \nu}
e^{i \vec q \vec r}; \
\Psi^c_{\nu} (\vec r) = \int \frac{d \vec q}{(2\pi)^{3/2}} \sum_{s} V^*_{q s} v_{\vec{q} s \nu} e^{i \vec q \vec
r},
\end{eqnarray}
\noindent we have found

\begin{eqnarray}
\label{10} E_1(0) = \int d \vec r \{ \Psi^+ (\vec r) [(-i\vec \alpha \vec \nabla + \beta m_0) +
\frac{1}{2} \varphi (\vec r) ] \Psi (\vec r) - \nonumber\\
- \Psi^{+c} (\vec r) [(-i\vec \alpha \vec \nabla + \beta m_0) +
+ \frac{1}{2}e_0 \varphi (\vec r) ] \Psi^c (\vec r); \nonumber\\
\int {d \vec{r}} [\Psi^{+} (\vec r) \Psi (\vec r) + \Psi^{+c} (\vec r') \Psi^{c} (\vec r')] = 1;
\end{eqnarray}

\begin{eqnarray}
\label{11} \varphi (\vec r) = \alpha_0 \int \frac {d \vec{r}'} {|\vec r - \vec{r}'|} [\Psi^{+} (\vec
r') \Psi (\vec r') - \Psi^{+c} (\vec r') \Psi^{c} (\vec r')].
\end{eqnarray}

Extremum of this functional for the state with orbital momentum $l=0$ has led  to the system of equations in the dimensionless variables and  functions:
\begin{eqnarray}
\label{12} x = r m_0; \quad E = \epsilon m_0; \quad  e_0 \varphi(r) = m_0\phi(x);
\quad \frac{e^2_0}{4 \pi} = \alpha_0;\nonumber\\
u(x) \sqrt{m_0} = r g(r);\quad v(x)\sqrt{m_0} = r f(r);\quad u_1(x)\sqrt{m_0} = r g_1(r);\quad v_1(x)\sqrt{m_0} =
r f_1(r).
\end{eqnarray}

\begin{eqnarray}
\label{13} \frac{d u}{dx} - \frac{1}{x}u - (  1 - \phi(x)) v  = 0; \
\frac{d v}{dx} + \frac{1}{x}v - (  1 + \phi(x)) u = 0;
\nonumber\\
\frac{d u_1}{dx} + \frac{1}{x}u_1 - (  1 + \phi(x)) v_1 = 0; \
\frac{d v_1}{dx} - \frac{1}{x}v_1 - (\  1 - \phi(x)) u_1 = 0;
\nonumber\\
\phi(x) = \alpha_0 [ \int_{x}^{\infty} dy \frac{\rho(y)}{y} +
\frac{1}{x} \int_{0}^{x} dy \rho(y)];\nonumber\\
\rho(x) = [u^2(x) + v^2(x) - u^2_1(x) - v^2_1(x)].
\end{eqnarray}

Solution of this equations exists when the parameter

\begin{eqnarray}
\label{14} a = \alpha_0 \frac{1-C}{1+C} = a_0 \approx - 3.531... \
\int_0^{\infty} d x (u^2 + v^2) = \frac{1}{1+ C}; \ \int_0^{\infty} d x (u_1^2 + v_1^2) = \frac{C}{1+ C}
\end{eqnarray}

Taking into account the connection (\ref{7}) between the "bare" and "physical" charges    the   relation (\ref{1}) between the "bare" coupling
constant   $\alpha_0 $ and the observed value of the fine structure constant   $\alpha  $ has been found.

When the analogous calculations have been fulfilled for the state corresponding to the "physical" electron total momentum $\vec P$ the spectrum of the one particle excitation has been found:

\begin{eqnarray}
\label{15} E_1(P) = \sqrt{P^2 + m^2}; \quad m = \frac{m_0 \alpha T}{|a_0|}.
\end{eqnarray}

In this paper we consider by the same way the vacuum energy $E_0$ with the trial state vector $|\Phi_0 > \equiv | 0; 0; 0 >$. The result  includes three terms \cite{Bjorken}:

1) The Dirac field zero energy:

\begin{eqnarray}
\label{16}
<0| \hat H_D |0> = - 2 \sum_{\vec p} \sqrt{p^2 + m_0^2}   = - 2 \frac{V}{(2\pi)^3}\int  \sqrt{p^2 + m_0^2} d\vec p
\end{eqnarray}

2) Electromagnetic field vacuum energy:

\begin{eqnarray}
\label{17}
<0| \hat H_E |0> =  \sum_{\vec k \lambda} \omega (\vec k) \frac{1}{2}  =  \frac{V}{(2\pi)^3}\int   d \vec k  \ k
\end{eqnarray}

3) Coulomb vacuum energy:

\begin{eqnarray}
\label{18}
<0| \hat H_C|0> =  \frac{\alpha_0}{2}   \int \frac{d \vec r d \vec r'}{|\vec r - \vec r'|}<0| \rho (\vec r ) \rho (\vec r' ) |0>
\end{eqnarray}

It is important that there are one negative and two positive terms and all of them are equal to infinity. From the mathematical point of view it means that we deal with indefinite limit for $E_0$. In order to investigate this limit let us introduce a formal regularization function for integration in the momentum space

\begin{eqnarray}
\label{19}
\int d\vec p  \Rightarrow \int d\vec p f(\frac{p}{L}); \quad f(0) = 1; \quad f(\infty) = 0,
\end{eqnarray}
\noindent with $L$ as the regularization parameter ($ L \rightarrow \infty $).

Then these terms can be written as follows

\begin{eqnarray}
\label{20}
<0| \hat H_D |0> = - 2 \frac{V}{(2\pi)^3} \int d\vec p f(\frac{p}{L}) \sqrt{p^2 + m_0^2} = - 2 \frac{V}{(2\pi)^3}L^4 \int d\vec u f(u) \sqrt{u^2 + \frac{m_0^2}{L^2}},
\end{eqnarray}

\begin{eqnarray}
\label{21}
<0| \hat H_E |0> =     \frac{V}{(2\pi)^3} L^4 \int   d \vec u f(u) u.
\end{eqnarray}

And the last term can be transformed to the momentum space:

\begin{eqnarray}
\label{22}
<0| \hat H_C|0> = 2 \alpha_0 \int  \int  d \vec r d \vec r' \int \frac{ d \vec q}{2 \pi^2 q^2}\int  \int \frac{d \vec p d \vec p'}{(2 \pi)^6}f(\frac{p}{L}) f(\frac{p'}{L})e^{i (\vec p + \vec p' + \vec q) (\vec r - \vec r')} = \nonumber\\
= 2 \alpha_0 V  \int \frac{ d \vec q}{2 \pi^2 q^2}\int  \int \frac{d \vec p d \vec p'}{(2 \pi)^3}f(\frac{p}{L}) f(\frac{p'}{L})\delta (\vec p + \vec p' + \vec q) =  \nonumber\\ =  2 \alpha_0 V  \int \frac{ d \vec q}{2 \pi^2 q^2} \int \frac{d \vec p  }{(2 \pi)^3}f(\frac{p}{L}) f(\frac{|\vec p + \vec q|}{L})  = \nonumber\\
= 2 \alpha_0 V L^4 \int \frac{ d \vec u}{2 \pi^2 u^2}  \int \frac{d \vec v  }{(2 \pi)^3}f(v) f( |\vec v + \vec u| ).
\end{eqnarray}

Here $\vec u, \vec v$ are the dimensionless variables.

One can see that there is a possibility to turn the vacuum energy equal to zero if the "bare coupling constant" is chosen as follows:

\begin{eqnarray}
\label{23}
\alpha_0 (\frac{m_0}{L}) =  2 \pi^2 \frac{\int   d \vec u f(u) [2\sqrt{u^2 + m_0^2/L^2} - u]}{\int \frac{ d \vec u}{  u^2}  \int  d \vec v  f(v) f( |\vec v + \vec u| )} \approx \tilde{\alpha}_0 + O [\frac{m_0^2}{L^2}]; \nonumber\\
L \rightarrow \infty \quad \tilde{\alpha}_0  =   \pi^2 \frac{\int   d \vec u f(u)   u }{\int \frac{ d \vec u}{  u^2}  \int  d \vec v  f(v) f( |\vec v + \vec u| )} \equiv  I[f].
\end{eqnarray}

\section{Calculation of $\tilde{\alpha}_0$.}

A concrete form for the dimensionless regularization (cut-off)  function with the boundary condition (\ref{19}) is still not chosen. We suggest to choose it such a way that the value $\tilde{\alpha}_0$ in (\ref{23}) defined by the functional $I[f]$ depends on this form as less as possible. One can write the general form of this function  as  the series with $N$ indefinite coefficients:

\begin{eqnarray}
\label{24}
f(u) = e^{-u^2}\sum_{l = 0}^{N} C_l u^l; \quad C_0 = 1.
\end{eqnarray}

After integration over the angles the  value  $\tilde{\alpha}_0$ can be represented as follows:

\begin{eqnarray}
\label{25}
\int   d \vec u f(u)   u = 4 \pi \int_0^{\infty} du f(u) u^3  \equiv 4 \pi J_1; \nonumber\\
\int \frac{ d \vec u}{u^2}  \int  d \vec v   f(v) f( |\vec v + \vec u| ) =  \int \frac{ d \vec u}{  (\vec u - \vec v)^2}  \int  d \vec v   f(v) f(  u  ) = \nonumber\\
4 \pi^2 \int_0^{\infty} d u  \int_0^{\infty} d v u v     f(v) f(  u  )\ln {\frac{(u+v)^2}{(u - v)^2 }} \equiv 4 \pi^2 J_2; \nonumber\\
\tilde{\alpha}_0 = \pi \frac{J_1}{J_2}.
\end{eqnarray}

Let us now define two sets of integrals: the vector set

\begin{eqnarray}
\label{26}
I_l =   \int_{0}^{\infty} d u   e^{-u^2} u^{l + 3} ;
\end{eqnarray}

\noindent and the tensor set

\begin{eqnarray}
\label{27}
M_{k,r} =    \int_{0}^{\infty} d u  \int_{0}^{\infty} d v  u v    e^{-u^2 - v^2}  u^{k} v^{r} \ln {\frac{(u+v)^2}{(u - v)^2 + \delta^2 }} ;
\end{eqnarray}
\noindent where all indexes changes as  $0,1,...N$.

Then the  coupling constant should be calculated as:

\begin{eqnarray}
\label{28}
\tilde{\alpha}_0 =  \pi \frac{I_0 + A}{M_{0,0} + 2 A_1 + B}\nonumber\\
A = \sum_{l\neq 0}^N C_l I_l ; \
A_1 = \sum_{l\neq 0}^N C_l M_{0,l}; \
B = \sum_{k \neq 0 }^N \sum_{r \neq 0}^N M_{k,r} C_{k} C_{r}.
\end{eqnarray}

Minimal sensitivity of the calculated coupling constant from the coefficients $C_l$ on the considered class of functions corresponds to the conditions:

\begin{eqnarray}
\label{29}
\frac{\partial \tilde{\alpha}_0}{\partial C_l} = 0; \quad l \neq 0, \nonumber\\
I_l (M_{0,0} + 2 A_1 + B) = 2 (I_0 + A)(   M_{0,l} +   \sum_{k \neq 0 } M_{l,k}C_{k}).
\end{eqnarray}

Solutions of this system of equations can be found in general form:

\begin{eqnarray}
\label{30}
C_l =  K \sum_{k \neq 0 }^N (M^{-1})_{l,k} I_{k}  -    \sum_{k \neq 0 }^N  (M^{-1})_{l,k} M_{0,k}; \quad K = \frac{(M_{0,0} + 2 A_1 + B)}{2 (I_0 + A)} .
\end{eqnarray}

These solutions will be consistent if we use them in the definitions (\ref{28}) of the parameters  $A, A_1, B$

\begin{eqnarray}
\label{31}
A = K \sum_{k \neq 0 }\sum_{l \neq 0}I_l (M^{-1})_{l,k} I_k  - \sum_{k \neq 0 }\sum_{l \neq 0}I_l (M^{-1})_{l,k} M_{0,k}   \equiv K T - T_1; \nonumber\\
A_1 = K \sum_{l\neq 0}\sum_{l \neq 0}  M_{0,l}(M^{-1})_{l,k}   I_{k} - \sum_{l\neq 0}\sum_{l \neq 0}  M_{0,l}(M^{-1})_{l,k}   M_{0,k}
\equiv K T_1 - T_2; \nonumber\\
B =  K^2 \sum_{k \neq 0 }\sum_{r \neq 0} \sum_{l \neq 0 }\sum_{m \neq 0}M_{k,r} (M^{-1})_{l,k}   I_{l} (M^{-1})_{r,m}   I_{m} -  \nonumber\\
- 2 K \sum_{r \neq 0} \sum_{l \neq 0 }\sum_{m \neq 0}M_{k,r} (M^{-1})_{l,k}  M_{0,l} (M^{-1})_{r,m}   I_{m} + \nonumber\\ +\sum_{k \neq 0 }\sum_{r \neq 0} \sum_{l \neq 0 }\sum_{m \neq 0}M_{k,r} (M^{-1})_{l,k}   M_{0,l} (M^{-1})_{r,m}  M_{0,m}  =  K^2 T  - 2 K T_1 + T_2; \nonumber\\
T_2 =   \sum_{k \neq 0 }\sum_{l \neq 0}  M_{0,k} (M^{-1})_{l,k}   M_{0,l}.
\end{eqnarray}

This leads the following system of the algebraic equations:

\begin{eqnarray}
\label{32}
2 (A + T_1)(I_0 + A) - [(M_{0,0} + 2 A_1 + B)] T = 0;   \nonumber\\
(A + T_1)T_1 - (A_1 + T_2) T = 0;    \nonumber\\
  (B - T_2)T  -  (A + T_1)^2  +   (A_1 + T_2)T = B T - (A + T_1)^2 + A_1 T = 0.
\end{eqnarray}

If one  introduces new unknown variable $(A + T_1) = x$ the closed equation and its solutions can be found:

\begin{eqnarray}
\label{33}
x^2 + 2 x (I_0 - T_1  )   - (M_{0,0} - T_2)T  = 0 \nonumber\\
x_{1,2}  = - (I_0 - T_1  ) \pm \sqrt{(I_0 - T_1  )^2 + (M_{0,0} - T_2)T}.
\end{eqnarray}

It results in:

\begin{eqnarray}
\label{34}
 A    = x - T_1; \
A_1 = - T_2 + \frac{x T_1 }{T}; \
B = \frac{x^2}{T} +  T_2 -  \frac{x T_1}{T}.
\end{eqnarray}

\begin{eqnarray}
\label{35}
\tilde{\alpha}_0 = \pi \frac{I_0 + A}{M_{0,0} + 2 A_1 + B}
\end{eqnarray}

\section{Numerical results and discussion}

It is evident that the numerical value $\tilde{\alpha}_0$ depends on the number $N$ in the series (\ref{24}) which defines the dimension of the matrixes in (\ref{31}). We have found numerically that for the fixed $N$ the first root of the equation (\ref{33}) corresponds to minimal value $\tilde{\alpha}_0$ and the second one corresponds to maximal $\tilde{\alpha}_0$. Both values become closer to each other with increasing of $N$. It proved that the positive solution $\tilde{\alpha}_0 > 0$ could be found for $N \leq 12$ and the maximal value of $\tilde{\alpha}_0 $ has corresponded to $N =12$. In this case the following values have been calculated:

\begin{eqnarray}
\label{36}
I_0 = 0.5000; \ M_0 = 0.768306; \ T = 29.1837; \ T_1 = 0.500496; \ T_2 = 0.768195; \nonumber\\
\tilde{\alpha}_0 \approx 239.21.
\end{eqnarray}

One can see that the condition $\tilde{\alpha}_0 \gg 1$ is in a qualitative agreement with the value $\alpha_0$ from (\ref{1}). It is not surprisingly that these parameters do not coincide exactly. In the present analysis the vacuum energy was calculated for only one fermion and one boson fields referred to QED. Let us suppose that there are additionally the set of quantum fields including $N_F^{(c)}$ charge and $N_F^{(0)}$ neutral fermion fields with degeneracy $g_F^{(c)}$ and $g_F^{(0)}$ correspondingly. Analogous values $N_B^{(c)}, g_B^{(c)}$, $N_B^{(0)}, g_B^{(0)}$   could be defined for neutral boson fields. If the vacuum energy of all these fields are taken into account by the same way as for QED the value $\tilde{\alpha}_0 $ could be recalculated as follows:

\begin{eqnarray}
\label{37}
\tilde{\alpha}_0 \Rightarrow \tilde{\alpha}_0 \frac{ 1 + g_F^{(c)}N_F^{(c)} + g_F^{(0)} N_F^{(0)} - 1/2(g_B^{(c)}N_B^{(c)} + g_B^{(0)} N_B^{(0)}) }{1 + 1/4([g_F^{(c)}]^2 N_F^{(c)}  + [g_B^{(c)}]^2 N_B^{(c)})}.
\end{eqnarray}

As for example, if one includes the contribution of vacuum energy from neutrino ($\nu_e$) the calculated value becomes

\begin{eqnarray}
\label{38}
\tilde{\alpha}_0 \approx 717.63,
\end{eqnarray}
\noindent that is much closer to the value (\ref{1}).

Thus, it is shown in the paper that two independent ways for definition of the "bare" coupling constant between electromagnetic and matter fields are qualitatively agreed each other. It can be considered as the basis for further analysis of this problem.

\section{Acknowledgments}

Authors are very grateful to Professor J.Bjorken for stimulating discussion.

\end{large}

\end{document}